\documentclass{article}
\usepackage{authblk}
\usepackage[utf8]{inputenc}
\usepackage{geometry}
\usepackage{graphicx}
\usepackage{cite}
\usepackage{amsmath}
\usepackage{subfigure}
\usepackage{color}

\usepackage{latexsym}

\usepackage{xspace}

\begin{document}
\title{A Monolithic Integrated Microwave Photonics Filter}

\author[1]{Javier S. Fandi\~no}
\author[1,2]{Pascual~Mu\~noz}
\author[2]{David~Dom\'enech}
\author[1]{Jos\'e Capmany}

\affil[1]{iTEAM Research Institute, Universitat Polit\`ecnica de Val\`encia, C/ Camino de Vera s/n, Valencia 46022, Spain. e-mail: jcapmany@iteam.upv.es}
\affil[2]{VLC Photonics S.L., C/ Camino de Vera s/n, Valencia 46022, Spain. e-mail: david.domenech@vlcphotonics.com}

\maketitle

\begin{abstract}
Meeting the ever increasing demand for transmission capacity in wireless
networks will require evolving towards higher regions in the
radiofrequency spectrum, reducing cell sizes as well as resorting to
more compact, agile and power efficient equipment at the base stations,
capable of smoothly interfacing the radio and fiber segments. Photonic
chips with fully functional microwave photonic systems are promising
candidates to achieve these targets. Over the last years, many
integrated microwave photonic chips have been reported in different
technologies. However, and to the best of our knowledge, none of them
have fully integrated all the required active and passive components.
Here, we report the first ever demonstration of a microwave photonics
tunable filter completely integrated in an Indium Phosphide chip and
packaged. The chip implements a reconfigurable RF- photonic filter, it
includes all the required elements, such as lasers, modulators and
photodetectors, and its response can be tuned by means of control
electric currents. This demonstration is a fundamental step towards the
feasibility of compact and fully programmable integrated microwave
photonic processors.
\end{abstract}

Emerging information technology scenarios, such as 5G mobile
communications and Internet of Things (IoT), will require a flexible,
scalable and future- proof solution capable for seamlessly interfacing
the wireless and fiber segments of
communication~networks [1,2,3].~Microwave~photonics~(MWP) [4,5],the~interdisciplin
ary approach that combines radiofrequency and photonic systems, is the
best positioned technology to achieve this target. A very relevant
example is 5G wireless communications, which targets an extremely
ambitious range of requirements including [6,7], a~1000-fold increase in
capacity, connectivity for over 1~billion users, strict latency control,
as well as network flexibility via agile software programming. These
objectives call for a paradigm shift in the access network to
incorporate smaller cells, exploit the millimeter wave regions of the
radiofrequency spectrum and implement massive multiple-input multiple-
output at the base stations (BTSs) [7]. The successful integration of the
wireless and fiber segments thus relies on the possibility of
implementing agile and reconfigurable MWP subsystems, featuring
broadband operation, as well as low space, weight and power consumption
metrics. The solution consists in resorting to integrated microwave
photonics (IMWP) [8,9] chips allocated either in the BTS and/or the central
office in combination with radio over fiber transmission in the fiber
segment connecting them [10,11]. The two fundamental issues to be solved in
IMWP are related respectively to technology and architecture. First,
there is the need to identify the best material platform where to
implement MWP chips. Second, whether it would be better to follow an
application specific photonic integrated circuit (ASPIC) approach, where
a specific architecture is employed to implement a specific
functionality, or to resort to a generic programmable architecture. IMWP
ASPICs with certain complexity have been reported to date mainly in four
material platforms: indium phosphide (InP) [12-14], Silicon-on-Insulator
(SOI) [15-21], silicon nitride (Si3N4) [22-26] and chalcogenide glass [27,28].
Several functionalities have been demonstrated with a variable degree of
photonic (20-60\%) integration, as shown in Table~1. A different
approach is based on generic processors [29,30], where a common
architecture implements different functionalities by suitable
programming. A recent paper [31] reported the design of a programmable
optical core inspired by the concept of electronic field programmable
gate arrays. This approach is based on an optical core composed
implemented by a 2D waveguide mesh where the connections between
waveguides are controlled by means of tunable Mach-Zehnder
interferometers (MZIs). Researchers fabricated a simplified version of
the processor composed of two mesh cells, using a commercial Si3N4
waveguide technology known as TriPleX [32]. The reported processor featured
a free spectral range of 14 GHz and is fully programmable. A band-pass
filter with a tunable centre frequency that spans two octaves
(1.6-6~GHz) and a reconfigurable band shape (including a notch filter
and a flat-top resonance) was demonstrated. A reconfigurable processor
implementing signal integration, differentiation and Hilbert
transformation has also been recently reported in InP technology [33]. To
the best of our knowledge however, none of the above contributions has
reported to date the integration of all the required active (sources,
modulators and detectors) and passive (splitters, optical filters and
waveguides) photonic components in a single chip, either monolithically
or following a hybrid approach. Here we report, , the design,
fabrication, packaging and experimental demonstration of the first
monolithic IMPW filter that integrates all these elements in the same
substrate. The chip implements a reconfigurable RF-photonic filter that
employs a tunable distributed Bragg reflector laser (DBR); a
single-sideband optical modulator; a tunable optical filter based on a
ring assisted Mach Zehnder interferometer (RAMZI)[34]; and an on-chip
optical detector. This demonstration constitutes a fundamental step
forward in the implementation of a fully integrated MWP filter, and
opens the path for compact and programmable MWP signal processors, where
the RAMZI filter will be replaced by a 2D reconfigurable mesh and
multiple functionalities will be implemented by suitable programming of
the mesh interconnections.

  {\par\centering
   \resizebox*{0.48\textwidth}{!}{\includegraphics*{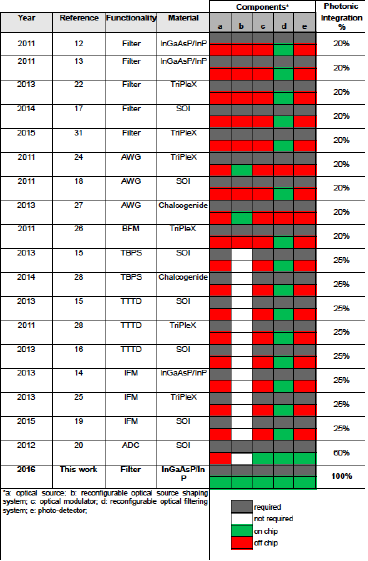}}
   
   Table 1: Overview of reported IMWP chips. AWG: Arbitrary waveform generation. BFM: Beamforming. TBPS: Tunable broadband phase shift. TTTD: Tunable true-time delay. IFM: Instantaneous frequency measurement. ADC: Photonic analog-to-digital conversion.
  }

\section*{Results} 
\subsection*{Basic concept}
Tunable MWP filtering is achieved in this
photonic integrated circuit (PIC) by exploiting the well-known mapping
between the optical and RF domains that happens when an optical
single-sideband modulation (SSB) is sent through a tunable optical
filter and is photodetected afterwards [35,36]. In brief, as shown in
Fig.~1a, the output RF tone at a given modulation frequency is produced
by the beating between the carrier and the optical SSB. As the
modulation frequency is swept, the optical filter modifies the amplitude
and phase of the sideband, while the carrier remains unchanged. This
means that, after photodetection, the RF response of the MWP filter is
just a scaled copy of the transfer function of the optical one, except
for a constant amplitude and phase factor (see Operation principle in
the Methods section). A MWP filter with a reconfigurable RF response can
thus be obtained by combining a tunable laser, an optical SSB modulator,
a photodetector and a tunable optical filter. The reconfigurability of
the response lies here on both the tunable laser, whose central
frequency can be changed to shift the whole RF transfer function, as
well as on the tunable filter, which can change both the central
frequency and the shape of the passband. Reconfigurable optical filters
can be implemented for example with integrated optical lattice
architectures [34,37]. Fig.~1b is a diagram of the designed InP PIC and its
different building blocks (BBs). First, light coming out of a tunable
distributed Bragg reflector laser (TL~1) is injected into a dual-drive
modulator (MZM), which is made up of a symmetric Mach-Zehnder
interferometer (MZI) with a couple of electro-optic phase shifters, one
on each arm. By introducing two RF signals with a 90$\deg$ relative phase
shift, and then setting the modulator at the quadrature bias point
(90$\deg$), an SSB modulation can be directly generated on-chip. This SSB
signal is then sent through a an optical filter that can be reconfigured
via thermo-optic heaters. The filter is based on the aforementioned
RAMZI architecture, where ring resonators of the same perimeter are
coupled to both branches of a symmetric Mach-Zehnder [38,39]. Because two
3~dB couplers are employed to split and recombine the signal in the
interferometer arms, note that this filter architecture has two inputs
and two outputs. One of the two available inputs is used to inject the
SSB signal coming from the modulator into the filter. Afterwards, the
modulated signal exists through one of the two outputs and is routed to
an on-chip photodetector (PD~4), where the optical to electrical mapping
finally takes place. The other free input of the RAMZI filter is routed
with an auxiliary waveguide (called Input) to an optical spot-size
converter located on one of the chip facets. The same is done for the
other free output port, which is also routed with an auxiliary waveguide
(called Output) to another spot- size converter on the same facet. The
chip was manufactured and packaged using an InP generic integration
technology on a multi-project wafer run [40] (see Fabrication and packaging
in the Methods section for more details). Figures~1c-1e show a 3D
representation of the PIC mask layout, a picture of a fabricated 6x4mm2
die, as well as a picture of one of the packaged chips, respectively.
Note that the final chip layout includes more components that the ones
described above, such as an extra tunable DBR laser laser, a phase
modulator and a tunable 2x3~MZI with three DC photodiodes located at its
output. These were employed when needed as auxiliary elements during
individual characterization of some of the BBs of the MWP filter, like
the tunable laser, the photodetector and the thermo-optic heaters.
Experimental procedures and results for these elements are given as
supplementary information (see Supplementary, Building block
characterization section). Because of its importance, the RAMZI filter
and the dual-drive MZM are described separately below.

\subsection*{RAMZI filter}
Integrated optical filters are typically made up of a
cascade (or lattice) of more basic, simpler optical elements, such as
ring resonators and Mach-Zehnder interferometers [36]. Depending on the
configuration and type of the employed basic unit cells, both finite
impulse response (FIR) and infinite impulse response (IIR) optical
filters can be implemented. In this case, a well-known architecture was
chosen based on a symmetric MZI loaded with ring resonators, also known
as RAMZI. A schematic diagram is shown in Fig.~2a. This architecture
allows for the implementation of optimum IIR bandpass filters, such as
the canonical Butterworth, Chebyshev and Elliptic approximations, while
featuring a fewer number of optical elements as compared with other
approaches [38]. In our case, the filter was designed as a fourth order
Chebyshev type~II filter, which ensures a flat passband and a sharper
roll-off at the expense of ripples in the stopband. Note that the
optical response is determined here by the coupling strength between the
ring resonators and the MZI branches, as well as by the relative optical
phase shifts of the rings and the MZI arms. The optical couplers that
connect the rings with the arms of the Mach- Zehnder are fixed, and were
implemented with custom 2x2~multimode interference couplers (MMIs)
featuring a tapered body [41]. The optical phases can however be changed by
means of thermo-optic heaters, which allow the tuning of the transfer
function. Like any other optical lattice structure, it has a periodic
frequency response, with a frequency periodicity commonly termed as Free
Spectral Range (FSR). This is determined by the perimeter of the ring
resonators and the group index of the waveguides, and it was designed to
be 20~GHz. Ideal (no loss) magnitude and phase responses, as well as the
simulated ones considering average propagation losses of 5.5~dB/cm, are
shown in Fig.~2b and 2c, respectively. In order to measure the tunable
RAMZI filter, the auxiliary Input and Output waveguides (see Fig.~1b and
1c) were employed. These are coupled to a pair of single-mode fibers
coming out of the metal package (Fig.~1e), which provide a mean for
interfacing the RAMZI filter with external equipment. Two difficulties
were nevertheless found during characterization of the filter. First,
these fibers are not polarization maintaining. Thus, when light from an
external source is injected into the chip, it is not generally possible
to prevent the excitation of both quasi-TE and quasi-TM modes (dominant
electric field component parallel/perpendicular to the chip surface,
respectively) in the PIC waveguides to which those fibers are coupled.
Since the chip was designed for TE polarization, care must be taken so
that a random mix of polarizations is not excited simultaneously. To
solve the aforementioned problem, a polarization alignment procedure was
devised, which is described as supplementary information (see
Supplementary, Polarization alignment procedure section). Second,
different experiments showed that significant polarization rotation was
taking place in the chip waveguides. These are also discussed with
detail in the supplementary information (see Supplementary, RAMZI filter
characterization~section).

\subsection*{Dual-drive MZM}
The purpose of the dual-drive MZM is to generate an SSB
modulation. This requires the RF signals driving both arms to be 90$\deg$ out
of phase, while the modulator should be biased at the quadrature point.
The exact bias voltage that needs to be applied to the modulator for it
to be at quadrature is generally unknown. A usual procedure is to
connect the two RF inputs to a broadband 90$\deg$ RF hybrid, and then to
slowly tune the bias point while observing the output modulation
spectrum in an optical spectrum analyzer (OSA) for a fixed modulation
frequency. A bias voltage is finally chosen for which the ratio between
the powers of the lower and upper sidebands is maximum. This ratio is
called SSB suppression ($\tau$), and it is usually expressed in dB.

{\par\centering
   \resizebox*{0.96\textwidth}{!}{\includegraphics*{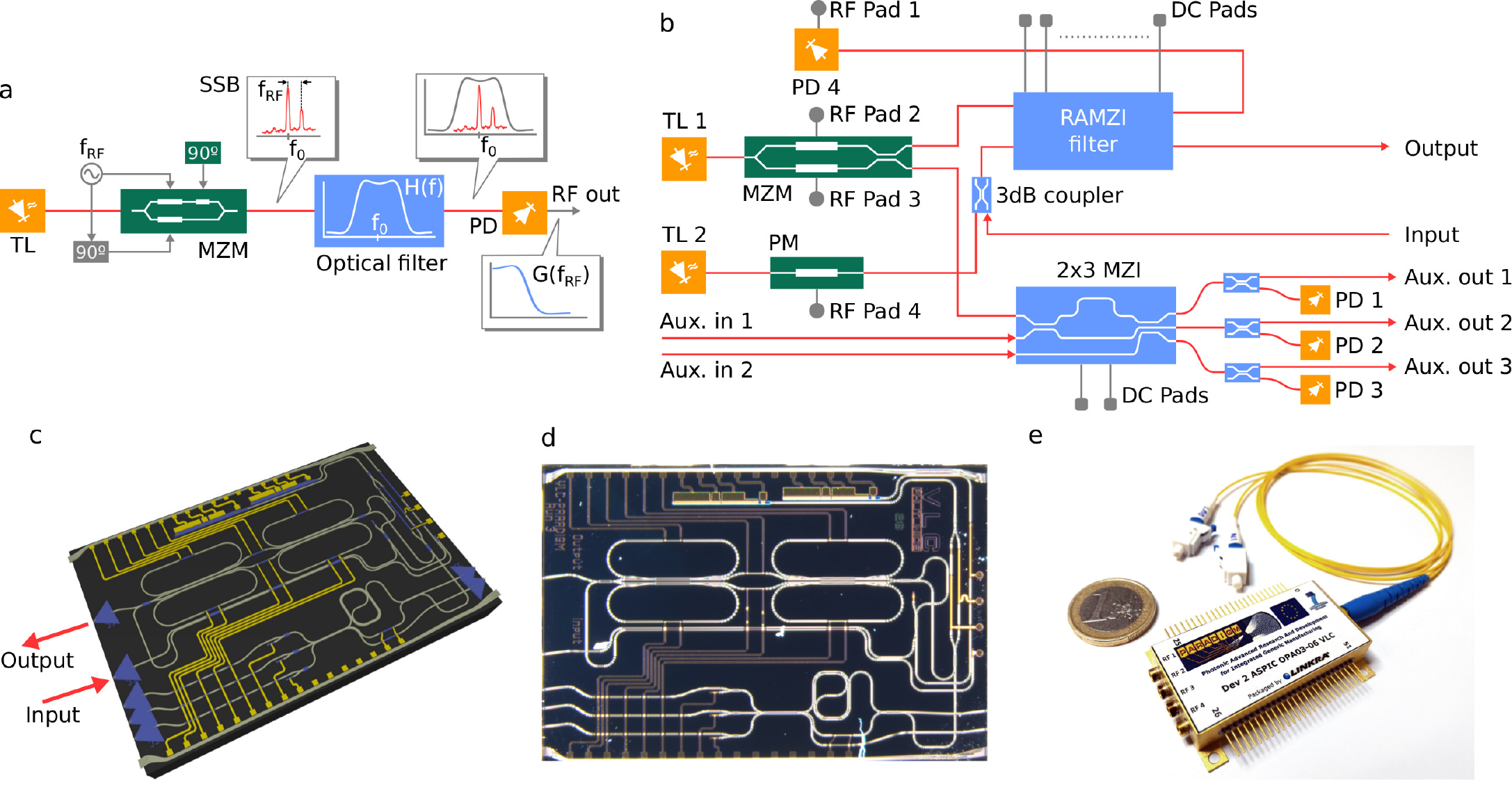}}
} 

Fig. 1. Operation principle and schematic diagram of the integrated MWP
filter. a, MWP filtering approach based on an optical SSB modulation and
a tunable optical filter. b, Schematic of the InP chip, showing all the
main building blocks. c, 3D representation of the chip layout. d,
Picture of a fabricated die (6x4 mm2). e, Packaged chip. TL: Tunable
laser. MZM: Dual- drive Mach-Zehnder modulator. PD: Photodetector. PM:
Phase modulator. MZI: Mach-Zehnder interferometer.

However, the layout of this PIC poses an extra difficulty: the optical
spectrum measured at the output fiber is not that of the modulator
itself, but that of the modulated signal after passing through the RAMZI
filter. As a consequence, the measured ratio does not match with the
real ratio at the output of the modulator. An alternative approach to
solve this issue was devised (see Adjustment of the dual-drive MZM in
the Methods section). The experimental setup for the MZM adjustment is
shown in Fig.~2d, while the final results are shown in Figs.~2e-2f.
Figure~2e shows the dependence of the median SSB suppression with bias
voltage. As it can be seen, it reaches a maximum of around 14~dB at
-8~V, which is the optimum bias point. This is better illustrated in the
inset, where an overlap of all the measured modulation spectra (solid
grey lines) is shown for this optimum voltage. They have been
conveniently renormalized and shifted in wavelength, so that the
fluctuations of the SSB suppression around the average value could be
better appreciated. A black solid line represents the average of all
these responses. Finally, Fig.~2f shows the experimental results for the
same procedure, but this time keeping the optimum voltage (- 8~V) and
instead sweeping the modulation frequency with the same input RF power.
As it can be seen, the achieved suppression is not constant with
modulation frequency. For example, it goes below 5~dB for frequencies
between 9 and 11~GHz, while it also shows a couple of dips around 7 and
8~GHz. This behaviour is attributed to poor RF performance of the
modulator electrodes, due to impedance mismatch at the end of the RF
line. At high frequencies a great part of the RF power is bouncing back
through the transmission lines, creating standing wave patterns that are
significantly affecting its broadband modulation behaviour, and
introducing unwanted power and phase imbalances. 

{\par\centering
   \resizebox*{0.96\textwidth}{!}{\includegraphics*{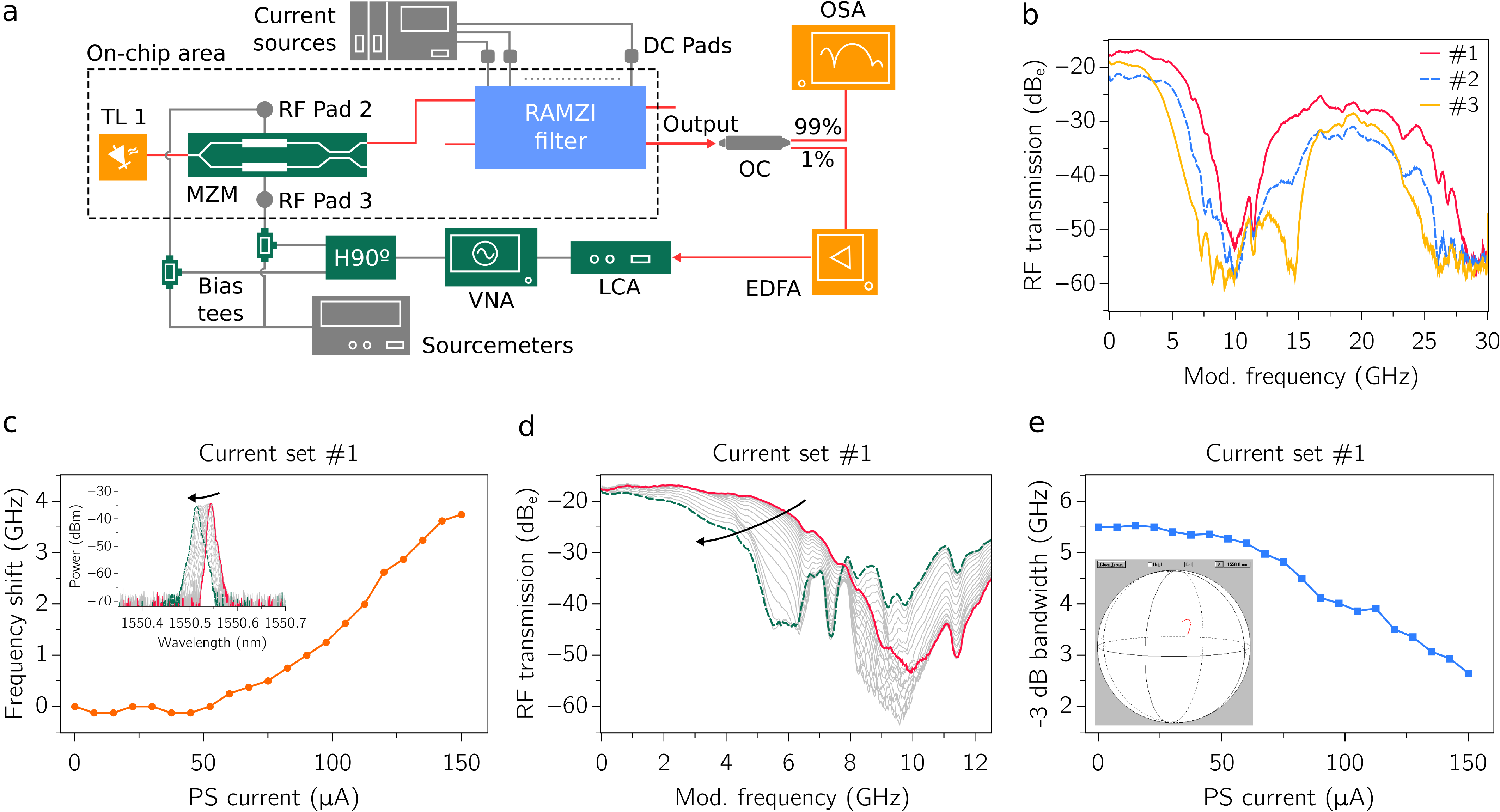}}
} 

Fig. 2.
RAMZI filter simulations and experimental adjustment of the dual- drive
MZM. a, Diagram of the RAMZI filter. b, Ideal (solid grey line)
magnitude response of the designed 4th order filter, and the simulated
one (dashed magenta line) when considering propagation losses of
5.5~dB/cm. c, Idem for the phase response. d, Experimental setup for
determining the optimum bias point of the dual-drive MZM. e, Dependence
of the measured median sideband suppression with the applied bias
voltage on one of the modulator arms, for a fixed modulation frequency
of 8.5~GHz. The grey lines in the inset shows all the captured
modulation spectra for the optimum bias point (-8~V), while the solid
magenta line is the average of all these traces. f, Idem, when the bias
is kept at the optimum point and the modulation frequency is swept
between 6 and 12 GHz.

\subsection*{Tunable MWP filter}
After the different BBs were characterized
individually, two experiments were carried out to demonstrate the basic
functionality of the MWP filter. The first one involved measuring its
electro/optical (E/O) response. That is, modulating the on-chip
dual-drive MZM and afterwards detecting the signal at the output of the
RAMZI filter with an external photodetector connected to one of the
fibers coming out of the PIC package. As discussed later in the text, a
tunable low-pass MWP filter has been obtained with this approach, by
first adjusting the transfer function of the RAMZI filter and then
tuning the central wavelength of the on-chip laser. In the second
experiment, a full electrical to electrical (E/E) measurement was
performed. That is, an electrical RF signal was again injected into the
on-chip dual-drive MZM, and the RF output at the integrated
photodetector (PD~4) was then measured. However, the presence of
significant electrical crosstalk was found to totally degrade the
response of the MWP filter. This is a very important lesson to be
learned when incorporating both driving and receiving RF electronics in
the same chip and is discussed with more detail in the Electrical to
Electrical (E/E) response of the MWP filter section of Supplementary. A
schematic diagram of the E/O experiment is shown in Fig.~3a. First, the
TL~1 is switched on, and a VNA and a 90$\deg$ RF hybrid are employed to
modulate the dual-drive MZM. This is polarized at the optimum bias point
(-8~V) to achieve the best SSB suppression. After the SSB modulation
goes through the RAMZI filter, it is collected at the fiber, which is
coupled to the auxiliary Output waveguide. Once the signal is outside
the chip, it is split in two distinct paths by a 99\%-1\% optical
coupler (OC). One one hand (1\%), it is injected into an OSA. On the
other hand (99\%), it is amplified by an erbium-doped fiber amplifier
(EDFA) and sent into a lightwave component analyzer (LCA, Agilent
N4373C). Due to the polarization rotation issues described in the RAMZI
filter characterization section (see Supplementary material), the
polarization of the optical signal at the input of the RAMZI filter when
coming from TL~1 is different to the one that results when an external
laser source is coupled to the Input waveguide. Since the filter
response is polarization dependent, the RAMZI filter will exhibit two
different transfer functions in these two situations. This ultimately
means that the response of the RAMZI filter as seen by the on-chip
optical SSB modulation cannot be adjusted by measuring the filter with
the external optical fibers, and must be optimized instead by directly
looking at its E/O transfer function. During the experiment, the
currents injected into the thermo-optic heaters were changed until a
response resembling the target filter shape was obtained. We found three
distinct sets of currents (\#1, \#2 and \#3) that yielded three
different, although similar, responses. These are plotted in Fig.~3b.
All of them feature a low-pass response with a -3~dB bandwidth of 5.5, 5
and 3~GHz, respectively. Besides, note that they also exhibit a resonant
peak near 20~GHz, which matches the target FSR of the filter.

{\par\centering
   \resizebox*{0.96\textwidth}{!}{\includegraphics*{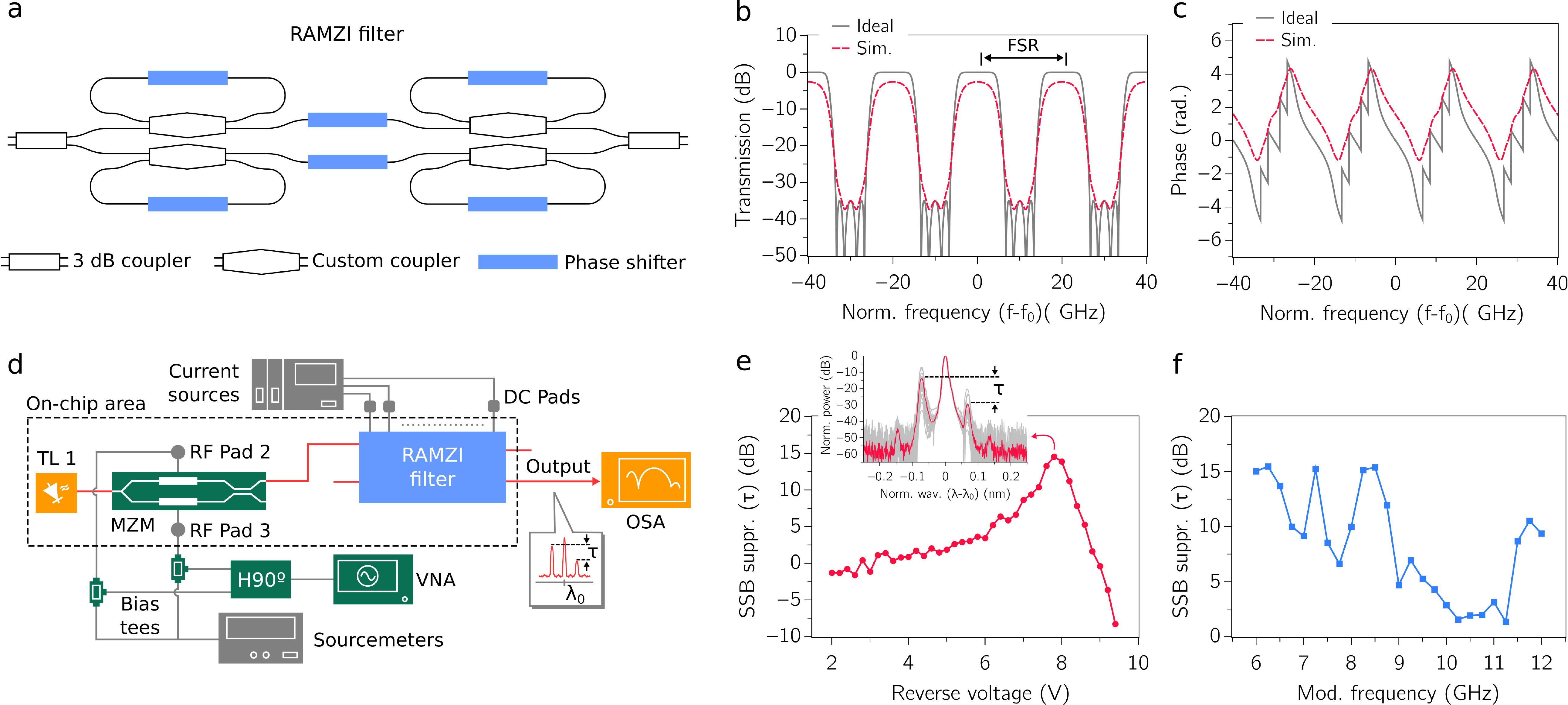}}
} 

Fig. 3. Experimental results for the tunable MWP filter (E/O). a,
Diagram of the experimental setup for the E/O measurement. b, Measured
E/O responses for three different current sets in the thermo-optic
heaters (\#1, \#2, \#3). c, Shift of TL~1 central frequency as a
function of the current injected into its phase shifting section (PS),
for the first current set (\#1). The inset shows the evolution of the
unmodulated emission spectrum as measured with an OSA. d, Evolution of
the measured E/O response as the frequency of TL~1 is shifted. e,
Computed dependence of the -3~dB bandwidth when current is injected into
the PS. The inset shows the measured evolution of the polarization state
on the Poincaré sphere at the output of the RAMZI filter.

To demonstrate the tunability of the MWP filter, several E/O responses
for one of these current sets (\#1) were stored as different currents
were applied to the PS shifting section of TL~1, which changed the
center frequency of the laser. The changes in laser frequency were
recorded by capturing the unmodulated output emission spectra in the
OSA. Results are shown in Figs.~3c-3e. Figure~3c represents the positive
shift in the central (optical) frequency of TL~1 (negative wavelength
shift) as the current was increased in the PS section. It was shifted
between 0 and 4~GHz for when the current was increased from 0 to 150~$\mu$A.
An overlapped representation of all the measured laser emission spectra
is shown in the inset, when no RF power was driving the modulator. For a
fixed optical frequency in TL~1, the different E/O responses obtained
during the (RF) frequency sweep are shown in Fig.~3d. The solid magenta
line represents the initial E/O response, when no current was injected
into the PS, while the dashed, turquoise green line is the response that
corresponds with the maximum injected current (150~$\mu$A). The MWP filter
exhibits tunable low-pass behaviour, where the width of the band is
reduced by increasing the laser frequency. The measured -3~dB bandwidth
as a function of the tuning current is plotted in Fig.~3e. The -3~dB
bandwidth decreases from a maximum of 5.5~GHz to a minimum of 2.5~GHz
(3~GHz range), which is similar to the increase in the frequency of the
laser (4~GHz) of Fig.~3c. The difference in these values, as well as the
observed evolution of the RF filter shape for different PS currents, can
be explained by both polarization rotation in the bends and the
frequency- dependent SSB suppression of the dual-drive MZM. Polarization
rotation is shown in the inset of Fig.~3e, where the measured
polarization state of the unmodulated light coming out of the PIC during
the frequency sweep is plotted in the Poincaré sphere. As the laser
frequency increases, the polarization state of the signal inside the PIC
changes, and so the effective transfer function of the RAMZI filter as
seen by the SSB modulation.

\section*{Discussion and summary}
In summary, we have reported on the design,
fabrication and experimental demonstration of a monolithically
integrated MWP filter in an InP PIC. The photonic packaged chip includes
a tunable laser, photodetectors, a dual- drive modulator, as well as a
RAMZI filter tunable via thermo-optic heaters. To the best of our
knowledge, this is the first fully integrated MWP filter reported to
date, and represents a significant step forward in the development of a
generic integrated MWP processor, where the specific RAMZI optical
filter will be replaced by a reconfigurable 2D mesh network. Integrated
MWP filters are of special interest because of their superior tuning
capabilities as compared to other RF technologies, and hold a
significant potential as the next generation of mobile wireless
standards move into higher frequency bands and require of wider
bandwidths. During the course of the measurements, several detrimental
effects, which will be fundamental in the process of implementing future
integrated MWP processors, were identified and studied in more detail.
In particular, a significant polarization rotation was observed in the
deeply etched waveguides of the InP platform, which was later confirmed
by means of simulations. This prevented the external measurement of the
RAMZI filter, which could not be controlled and tuned to exactly match
the original design target. Besides, severe RF crosstalk was observed
when performing full electrical to electrical (E/E) measurements of the
MWP filter. Nevertheless, the targeted tuning functionality was
demonstrated using an electrical to optical (E/O) approach, where an
auxiliary waveguide was employed to inject the SSB modulation at the
output of the RAMZI filter into an external photodetector. A
reconfigurable low-pass filter has been demonstrated this way, with a
-3~dB bandwidth tunable by 4~GHz when changing the frequency of the
on-chip tunable laser (TL~1). The hurdles found during the
characterization stage are not intrinsic to the proposed technique, but
rather due to fabrication imperfections and design issues, which can be
well solved by a proper PIC redesign and careful dedicated
manufacturing. Polarization rotation can be solved for example by either
reducing the sidewall angle of the waveguides or using a much higher
bending radius. Besides, the RF pads should be isolated and spaced as
much as possible to prevent RF crosstalk. Finally, a proper termination
of the dual-drive MZM arms with a right load would be beneficial to
increase the performance and broadband behaviour of the SSB modulation.

\section*{Methods}
\subsection*{Operation principle}
Assume that a laser with an average power
P0 and center frequency f0 is injected into an SSB optical modulator,
which is in turn being driven by an RF tone. The tone has the following
form: 

\begin{equation}
V_{RF}(t) = A\cos\left(2\pi f_{RF} t\right)
\end{equation}

where A is the RF voltage amplitude and fRF is the
modulation frequency. The complex low-pass equivalent of the optical
signal at the output of the modulator can be expressed, neglecting
optical losses, as follows

\begin{equation}
\tilde{E}_{SSB}\left(t\right) = \frac{\sqrt{P_o}}{2}\left[e^{j\pi V_{RF}(t) = A\cos\left(2\pi f_{RF} t\right)/V_{\pi}}+je^{j\pi V_{RF}(t) = A\sin\left(2\pi f_{RF} t\right)/V_{\pi}}\right]
\end{equation}

Here, V stands for the $\pi$
voltage of each phase shifter, and we have also assumed that a 90$\deg$ RF
hybrid is introducing a perfect $\pi$/2 phase shift between the two RF
signals that drive both arms of the SSB modulator. The modulator is also
being biased at the quadrature point (90$\deg$).

Usually, A is much lower than
V, so the small-angle approximation holds. That is, $\exp{jx}=(1-x^2/2)+jx$ for x $\ll$ 1. For the sake of simplicity,
second-order powers of the input signal are also neglected. Equation 2
can then be finally expressed as 

\begin{equation}
\tilde{E}_{SSB}\left(t\right) = \frac{\sqrt{P_o}}{2}\left[(1+j)+j\frac{\pi}{V_{\pi}}Ae^{j2\pi f_{RF} t}\right]
\end{equation}

The Fourier transform of this waveform leads to

\begin{equation}
\tilde{E}_{SSB}\left(f\right) = \frac{\sqrt{P_o}}{2}\left[(1+j)\delta (f)+j\frac{\pi}{V_{\pi}}A \delta (f-f_{RF})\right]
\end{equation}

which corresponds with a
single-sideband modulation. If this signal is introduced into an optical
filter, then at the output we get 

\begin{equation}
\tilde{E}_{out}\left(f\right) = \frac{\sqrt{P_o}}{2}\left[(1+j)H(0)\delta (f)+j\frac{\pi}{V_{\pi}}A H(f_{RF})\delta (f-f_{RF})\right]
\end{equation}
where H(f) stands for
the complex low-pass equivalent frequency response of the optical
filter. That is, H(0) means that the filter is evaluated at the
operating wavelength of the laser ($f_0$, or $\lambda_0$ in the wavelength domain).
The previous equation can be converted back to the time domain as

\begin{equation}
\tilde{E}_{out}\left(t\right) = \frac{\sqrt{P_o}}{2}\left[(1+j)H(0)+j\frac{\pi}{V_{\pi}}A H(f_{RF})e^{j2\pi f_{RF} t}\right]
\end{equation}
Finally, after detection, a time-varying photocurrent is
obtained that is proportional to the modulus squared of {[}pic{]}.
Neglecting the DC terms, and rewriting H(0) as $\kappa$, the RF component is
given by 

\begin{equation}
i(t) \propto |\kappa||H(f_{RF}| \cos(2\pi f_{RF} t + \angle H(f_{RF}) + \angle \kappa)
\end{equation}
Finally
Equation 7 means that the RF response of the MWP
filter (G(fRF)) is a scaled copy of the complex, low-pass equivalent of
the optical filter transfer function (H(f)), multiplied by a complex
constant ($\kappa$). That is 

\begin{equation}
G(f_{RF}) \propto \kappa^*H(f_{RF})
\end{equation}

As a consequence, a direct mapping
between the optical and RF domains is obtained.

\subsection*{Device fabrication and packaging}
The designed PIC was manufactured
within the european FP7 project PARADIGM (FP7-ICT-2009-5/257210). The
PARADIGM initiative aimed at facilitating access to state-of-the-art InP
foundries following a generic approach. That is, by providing external
users with a set of predefined BBs that are not trivial to design, and
then sharing the cost of PIC fabrication via multi-project wafer runs.
In this case, the chip was manufactured in a non-insulating InP platform
offered by the company Oclaro (United Kingdom). Deeply etched waveguides
are made of an n- doped InP substrate over which an InGaAsP
multi-quantum well structure is grown, forming the waveguide core.
Afterwards, the core region is covered with a layer of p-doped InP.
Finally, the whole structure is etched by about 3.6~$\mu$m from the top of
the p-doped region. The deep etch results in a high mode confinement,
which allows for a tight bending radius of 150~$\mu$m. In this case, a
waveguide width of 1.5~$\mu$m was employed. According to the available data,
it should experience propagation losses of about 5.5~dB/cm. Angled
spot-size converters with both vertical and lateral tapers were also
provided, featuring estimated coupling losses below 1~dB. The PIC was
packaged to facilitate its characterization. This service was offered by
some partners of the PARADIGM consortium, and in this case was done by
the company Linkra (Italy). The PIC layout was designed to meet the
requirements imposed by the standard packaging process. This sets a
maximum number of optical inputs/outputs, DC and RF pads; as well as it
specifies a predefined location for all of them (see Figs.~1c-1e). A
maximum of 48 DC pads can be employed, which are all located on the
upper/lower facets of the die. Additionally, up to 4 RF pads are
available to the designer, all placed close to the right chip facet.
Besides, up to two standard single-mode fibers can be coupled to on-chip
spot-size converters, which are located on the left chip facet. Inside
the metallic enclosure, the RF pads are wire-bonded to a ceramic
substrate that routes the RF signals to external GPPO connectors (DC to
65~GHz). The DC pads are also wire-bonded and then connected to a set of
equally spaced pins in the external metallic case. A thermistor and a
Peltier cell are also included in the package to provide accurate
temperature control.

\subsection*{Adjustment of the dual-drive MZM}
The experimental setup employed to
find the optimum bias point of the MZM is shown in Fig.~2d. It can be
described as follows. A microwave vector network analyzer (VNA, Agilent
PNA-X, N5245A, 10~MHz-50~GHz) first generates an RF tone with constant
amplitude and frequency. This tone is injected into a 90º RF hybrid
(Marki microwave, QH0R714, 0.7-14.5~GHz), whose two outputs are then
connected to a couple of broadband bias tees (SHF Communications,
SHF-BT-45-D, 20~kHz-45~GHz). In turn, they are used to drive the two
ports of the dual-drive MZM (RF pads 2 and 3 in Fig.~1b). Now, for each
reverse bias voltage in one of the MZM arms, we apply a different set of
independent, uniformly distributed random currents in the thermo-optic
heaters of the RAMZI filter, and finally capture the modulation spectrum
exiting through the output fiber. The power ratio between the lower and
upper sidebands ($\tau$) is measured and stored for each current set. It is
expected that the measured ratio will fluctuate up and down due to the
influence of the RAMZI filter, which is changing for each set of random
currents. However, if a sufficiently high number of modulation spectra
are measured, then the median ratio extracted from these measurements
will converge to the real value. That is, the real SSB suppression as it
would be measured at the output of the modulator. In the experiment, a
tone of +5~dBm and a frequency of 8.5~GHz was first used to modulate the
MZM. Besides, a constant reverse voltage of -2~V was applied in one of
the MZM arms, while the bias in the other arm was swept between 2 and
9.4~V in steps of 0.2~V. For each bias value, we applied a set of 25
different, independent, uniformly distributed random currents in the 6
thermo-optic heaters of the RAMZI filter, and the output spectrum was
captured in an OSA (Advantest~Q8384).

\section*{References}
\begin{enumerate} 
\item Novak, D. et al., Radio-Over-Fiber Technologies for Emerging
Wireless Systems. IEEE J. of Quantum Electron. 52, 1-11 (2016).
\item Waterhouse, R. \& Novak, D. Realizing 5G: Microwave Photonics for 5G
Mobile Wireless Systems. IEEE Microw. Mag. 16, 84-92 (2015). 
\item Technology
Focus on Microwave Photonics. Nature Photon. 5, 723 (2011). 
\item Capmany, J.
\& Novak, D. Microwave Photonics combines two worlds. Nature Photon. 1,
319-330 (2007). 
\item Yao, J. Microwave Photonics. J. Lightw. Technol. 27,
314-335 (2009). 
\item Andrews, J.G., Buzzi, S., Wan, C., Hanly, S.V., Lozano,
A., Soong, A.C.K. \& Zhang, J.C. What Will 5G Be? IEEE J. Sel. Areas
Commun. 32, 1065-1082 (2014). 
\item Gosh, Al. et al. Millimeter-Wave Enhanced
Local Area Systems: A High-Data- Rate Approach for Future Wireless
Networks. IEEE J. Sel. Areas Commun. 32, 1152-1163 (2014). 
\item Marpaung, D.
et al. Integrated Microwave Photonics. Laser Photon. Rev. 7, 506-538
(2013). 
\item Iezekiel, S., Burla, M., Klamkin, J., Marpaung, D. \& Capmany,
J. RF Engineering Meets Optoelectronics: Progress in Integrated
Microwave Photonics. IEEE Microw. Mag. 16, 28-45 (2015). 
\item Mitchell, J.E.
Integrated Wireless Backhaul Over Optical Access Networks. J. Lightw.
Technol. 32, 3373--3382 (2014). 
\item Liu, C., Wang, J., Cheng, L., Zhu, M. \&
Chang, G.-K. Key. Key Microwave- Photonics Technologies for
Next-Generation Cloud-Based Radio Access Networks. J. Lightw. Technol.
32, 3452--3460 (2014). 
\item Norberg, E.J., Guzzon, R.S., Parker, J.S.,
Johansson, L.A. \& Coldren, L.A. Programmable Photonic Microwave Filters
Monolithically Integrated in InP/InGaAsP. J. Lightw. Technol. 29,
1611-1619 (2011). 
\item Guzzon, R., Norberg, E., Parker, J., Johansson, L. \&
Coldren, L. Integrated InP-InGaAsP tunable coupled ring optical bandpass
filters with zero insertion loss. Opt. Express 19, 7816-7826 (2011).
\item Fandino, J.S. \& Munoz, P. Photonics-based microwave frequency
measurement using a double-sideband suppressed-carrier modulation and an
InP integrated ring-assisted Mach-Zehnder interferometer filter. Opt.
Lett. 38, 4316--4319 (2013).
\item Burla, M., Cortes, L.R., Li, M., Wang, X.,
Chrostwoski, L. \& Azaña, J. On- chip ultra-wideband microwave photonic
phase shifter and true time delay line based on a single phase-shifted
waveguide Bragg grating. in IEEE International Topical Meeting on
Microwave Photonics, 92-95 (2013). 
\item Shi, W., Veerasubramanian, V., Patel,
D. \& Plant, D. Tunable nanophotonic delay lines using linearly chirped
contradirectioinal couplers with uniform Bragg gratings. Opt. Lett. 39,
701-703 (2014). 
\item Guan, B. et al. CMOS Compatible Reconfigurable Silicon
Photonic Lattice Filters Using Cascaded Unit Cells for RF-Photonic
Processing. IEEE J. Sel. Topics Quantum Electron. 20, 359-368 (2014).
\item Khan, M.H., Shen, H., Xuan, Y., Zhao, L., Xiao, S., Leaird, D.E.,
Weiner, M.A. \& Qi, M. Ultrabroad-bandwidth arbitrary radiofrequency
waveform generation with a silicon photonic chip-based spectral shaper.
Nature Photon. 4, 117--122 (2010). 
\item Pagani, M. et al. Instantaneous
frequency measurement system using four- wave mixing in an ultra-compact
long silicon waveguide. in European Conference on Optical Communication
(ECOC), 1-3 (2015). 
\item Khilo, A.. et al. Photonic ADC: overcoming the
bottleneck of electronic jitter. Opt. Express 20, 4454-4469 (2012).
\item Wang, J. et al. Reconfigurable radio-frequency arbitrary waveforms
synthesized in a silicon photonic chip. Nat. Commun. 6, 5957 (2015).
\item Marpaung, D., Morrison, B., Pant, R., Roeloffzen, C., Leinse, A.,
Hoekman, M., Heideman, R. \& Eggleton, B.J. Si3N4 ring resonator-based
microwave photonic notch filter with an ultrahigh peak rejection. Opt.
Express 21, 23286--23294 (2013). 
\item Zhuang, L., Taddei, C., Hoekman, M.,
Leinse, A., Heideman, R., van Dijk, P. \& Roeloffzen, C. Ring
resonator-based on-chip modulation transformer for high-performance
phase-modulated microwave photonic links. Opt. Express 21, 25999--26013
(2013).
\item Marpaung, D., Chevalier, L., Burla, M. \& Roeloffzen, C. Impulse
radio ultrawideband pulse shaper based on a programmable photonic chip
frequency discriminator. Opt. Express 19, 24838--24848 (2011). 
\item Marpaung,
D. On-Chip Photonic-Assisted Instantaneous Microwave Frequency
Measurement System. IEEE Photon. Technol. Lett. 25, 837--840 (2013).
\item Burla, M., Marpaung, D., Zhuang, L., Roeloffzen, C., Khan, M.R., Leinse,
A., Hoekman, M. \& Heideman, R. On-chip CMOS compatible reconfigurable
optical delay line with separate carrier tuning for microwave photonic
signal processing. Opt. Express 19, 21475--21484 (2011). 
\item Tan, K. et al.
Photonic-chip-based all-optical ultra-wideband pulse generation via XPM
and birefringence in a chalcogenide waveguide. Opt. Express 21,
2003-2011 (2013).
\item Pagani, M. et al. Tunable wideband microwave photonic
phase shifter using on-chip stimulated Brillouin scattering. Opt.
Express 22, 28810-28818 (2014). 
\item Perez, D., Gasulla, I. \& Capmany, J.
Software-defined reconfigurable microwave photonics processor. Opt.
Express 23, 14640-14654 (2015). 
\item Capmany, J., Gasulla, I. \& Pérez, D.
Microwave photonics: The programmable processor. Nature Photon. 10, 6-8
(2016). 
\item Zhuang, L., Roeloffzen, C.G.H., Hoekman, M., Boller, K.-J. \&
Lowery, A.J. Programmable photonic signal processor chip for
radiofrequency applications. Optica 2, 854-859 (2015). 
\item Roeloffzen, C.G.
et al. Silicon nitride microwave photonic circuits. Opt. Express 21,
22937-22961 (2013). 
\item Liu, W. et al. A fully reconfigurable photonic
integrated signal processor. Nature Photon. 10, 190-195 (2016). 
\item Madsen,
C.K. \& Zhao, J. H., Optical Filter Design and Analysis: A Signal
Processing Approach. Wiley (1999). 
\item Roman, J., Frankel, M.Y. \& Esman,
R.D. Spectral characterization of fiber a gratings with high resolution.
Opt. Lett. 23, 939--941 (1998). 
\item Hernandez, R., Loayssa, A. \& Benito, D.
Optical vector network analysis based on single-sideband modulation.
Opt. Eng. 43, 2418--2421 (2004). 
\item Jinguji, K. \& Oguma, M. Optical
Half-Band Filters. J. Lightw. Technol. 18, 252--259 ( 2000). 
\item Madsen.
C.K. Efficient Architectures for Exactly Realizing Optical Filters with
Optimum Bandpass Designs. IEEE Photon. Technol. Lett. 10, 1136--1138
(1998). 
\item Madsen, C.K. General IIR Optical Filter Design for WDM
Applications Using All-Pass Filters. J. Lightw. Technol. 18, 860--868
(2000). 
\item M.K. Smit et al., ``An introduction to InP-based generic
integration technology,'' Semiconductor Science and Technology, vol.~29,
no 8, p.~083001 (2014). 
\item Besse, P.A., Gini, E., Bachmann, M. \& Melchior,
H. New 2x2 and 1x3 Multimode Interference Couplers with Free Selection
of Power Splitting Ratios. J. Lightw. Technol. 14, 2286--2293 (1996).
\item Halir, R., Vivien, L., Le Roux, X., Xu, D.-X. \& Cheben, P. Direct and
Sensitive Phase Readout for Integrated Waveguide Sensors. IEEE Photon.
J. 5, 6800906 (2013).
\item van Dam, C. et al. Novel Compact Polarization
Converters Based on Ultra Short Bends. IEEE Photon. Technol. Lett. 8,
1346--1348 (1996). 
\item Morichetti, F., Melloni, A. \& Martinelli, M. Effects
of Polarization Rotation in Optical Ring-Resonator-Based Devices. J.
Lightw. Technol. 24, 573--585 (2006).
\end{enumerate}

{\bf Acknowledgments.} The authors acknowledge financial support by the
Spanish CDTI NEOTEC start-up programme, the European Commission
FP7-PARADIGM project, the Generalitat Valenciana PROMETEO 2013/012
research excellency award, the Spanish MINECO project TEC2013-42332-P,
acronym PIF4ESP, project FEDER UPVOV 10-3E-492 and project FEDER UPVOV
08-3E-008. J.S. Fandi\~no acknowledges financial support through FPU grant
AP2010-1595.

{\bf Author contributions.} J.S.F. and D.D. designed the chip. J.S.F.
conceived the experiments and performed the measurements. J.S.F., P.M.
and J.C. analyzed the data and wrote the paper. P.M and J.C managed the
project.

{\bf Additional information.} The authors declare no competing financial
interests. Reprints and permission information is available online at
http://npg.nature.com/reprintsandpermissions/. Correspondence and
requests for materials should be addressed to J.C.

\end{document}